\documentclass{aa}  
\usepackage{natbib}
\bibpunct{(}{)}{;}{a}{}{,} 

\usepackage{newtxtext,newtxmath}
\usepackage[T1]{fontenc}
\usepackage{ae,aecompl}
\usepackage{comment}
\usepackage[hidelinks]{hyperref}
\usepackage{multirow}

\usepackage{graphicx}	
\usepackage[export]{adjustbox}

\usepackage{xcolor}

\def\gsim{\;\raise0.3ex\hbox{$>$\kern-0.75em\raise-1.1ex\hbox{$\sim$}}\;}

\begin{document}

\title{Iron He-triplet signatures of shocks in the hottest galaxy clusters}
\subtitle{$\mathtt{Z/W}$ line ratio, line broadening, and electron-ion temperature equilibration}

\author{
Eugene~Churazov \inst{1}
\and
Yuri~Ralchenko \inst{2,3,4}
\and
Ildar~I.~Khabibullin \inst{5,1} 
\and 
John~C.~Raymond \inst{6}
\and
Annie~Heinrich \inst{7}
\and
Irina~Zhuravleva \inst{7}
\and
Reinout~J.~van~Weeren \inst{8}
\and
Congyao~Zhang \inst{9,7}
}

\institute{
Max Planck Institute for Astrophysics, Karl-Schwarzschild-Str. 1, D-85741 Garching, Germany 
\and
Department of Astronomy, University of Maryland, College Park, MD 20742, USA
\and
X-ray Astrophysics Laboratory, NASA Goddard Space Flight Center, Greenbelt, MD 20771, USA
\and
Center for Research and Exploration in Space Science and Technology, NASA Goddard Space Flight Center, Greenbelt, MD 20771, USA
\and
Rudolf Peierls Centre for Theoretical Physics, Department of Physics, University of Oxford, Clarendon Laboratory, Parks Rd, Oxford, OX1 3PU, United Kingdom
\and
Center for Astrophysics | Harvard \& Smithsonian, 60 Garden St., Cambridge, MA 02138, USA
\and
Department of Astronomy \& Astrophysics, University of Chicago, Chicago, IL 60637, USA
\and
Leiden Observatory, Leiden University, PO Box 9513, 2300 RA Leiden, The Netherlands
\and
Department of Theoretical Physics and Astrophysics, Masaryk University, Brno 61137, Czechia
}

\abstract{
A merger of clusters naturally drives shocks with Mach number $\mathscr{M}\lesssim 3$ in the intra-cluster medium (ICM). This process creates several distinct signatures, including sharp surface brightness ``edges'', temperature, and gas velocity jumps. 
The low density of the ICM implies that the ionization balance and electron-ion equilibration times can be long enough to produce a set of additional observable signatures. Here, we focus on two ``transient'' spectral signatures accessible with the high-energy-resolution telescopes such as \textit{XRISM}, even for unfavorable geometry, e.g., when we are looking inside the Mach cone of the shock, precluding the appearance of sharp edges in X-ray images. In this work, we focus on (i) the $\mathtt{Z/W}$ line ratio of the Fe~XXV triplet and (ii) the contribution of ions with $T_i>T_{\rm e}$ to the line width, which might be mistakenly interpreted as the gas turbulence. We demonstrate that the $\mathtt{Z/W}$ ratio can serve as a proxy for the non-equilibrium state of the shocked ICM and facilitate interpretation of the line broadening. We conclude that these spectral signatures are within reach with missions like \textit{XRISM} and can be used to constrain the heating of electrons at the collisionless cluster shocks, as well as the rate of subsequent temperature equilibration between different particle species.}

\titlerunning{z/w ratio}

\keywords{Galaxies: clusters: intracluster medium -- X-rays: galaxies: clusters}
    
\maketitle

\section{Introduction}
\label{s:intro}
A merger of clusters is a common phenomenon seen in cosmological simulations and observations. Minor mergers (one of the merging clusters is much less massive than the other) are very frequent, while equal mass mergers are rarer. The latter often lead to spectacular features revealing the physics of the merger process \citep[see, e.g.][for a review]{2007PhR...443....1M}.       
The ICM temperature in galaxy clusters scales with the mass as $T_{X}\propto GM_{\rm v}/R_{\rm v}\sim V^2_{\rm v}$, where the subscript ``v'' stands for characteristic ``virial'' quantities.  Since the sound speed and the virial velocity follow the same scaling with mass, one expects a similar range of the shock Mach numbers, independent of the cluster mass.   
For a Navarro-Frenk-White potential with a concentration parameter of $c\sim 2-6$ \citep[e.g.,][]{2007MNRAS.379..190C}, the velocity of a point mass on a parabolic trajectory reaches $\sim 3.05-3.3 V_{\rm v}$ in the cluster center. This sets a natural characteristic value for the maximal Mach number, $\mathscr{M}\lesssim 3$, which one can expect during the pericenter passage. We discuss a generic case throughout the text, with examples such as the Bullet \citep{2002ApJ...567L..27M} and the Peanut \citep{2025A&A...693A..55L} clusters in mind. For illustration, we assume an initial gas temperature of $7\,{\rm keV}$ (for $M_{\rm v}\sim 10^{15}\,M_\odot$) and a shock with $\mathscr{M}=3.2$. 

In the context of cluster shocks, the ICM density can be very low, $n_e\sim 10^{-3}-10^{-4}\,{\rm cm^{-3}}$, implying that the relaxation time scales can be of order of $10^8\,{\rm yr}$, and departures from ionization equilibrium or different temperatures of electrons and ions might persist for a long enough time to produce observable signatures \citep[e.g.,][]{1997ApJ...491..459F,1998ApJ...495..630C,1999ApJ...520..514T,2007PhR...443....1M,2009ApJ...707.1141W,2010PASJ...62..335A,2012MNRAS.423..236R,2019A&A...628A.100D,2024ApJ...962..161S,2025ApJ...992...62N}. A similar set of processes defines the properties of the heated plasma in supernova remnants shocks in the interstellar medium (ISM) \citep[e.g.,][]{2008SSRv..134..141B,2012A&ARv..20...49V,2023ApJ...949...50R}. However, unlike the ISM, the ICM is hot (except for the very outskirts), and the sonic Mach number $\mathscr{M}$ is typically not much larger than $3$. Another difference is that the bulk of the ICM has the plasma $\beta=\frac{P_{\rm ICM}}{B^2/8\pi}\approx 100$, where $P_{\rm ICM}$ is the ICM thermal pressure and $B$ is the magnetic field, and, therefore, the Alfven Mach number is larger, $\mathscr{M}_{\rm A}=\beta^{1/2}\mathscr{M}\sim 10 \mathscr{M}$. 

The He-like triplet diagnostic is one of the powerful probes of low-density hot collision-dominated plasma \citep[e.g.,][]{2001A&A...376.1113P}. The most prominent lines are normally the well separated resonance $\mathtt{W}$ ($1s^2$  $^1S_0$ -- $1s2p$ $^1P_1$) line and the forbidden magnetic-dipole $\mathtt{Z}$ ($1s^2$  $^1S_0$ -- $1s2s$ $^3S_1$) line, while the intercombination $\mathtt{Y}$ ($1s^2$  $^1S_0$ -- $1s2p$ $^3P_1$) line and the magnetic-quadrupole $\mathtt{X}$ ($1s^2$  $^1S_0$ -- $1s2p$ $^3P_2$) line are often blended by dielectronic satellites.
The NEI signatures associated with the He-like triplet of high-Z elements have long been considered in the literature \citep[e.g.,][]{1978A&A....65..115M,1999LNP...520..189L,2004MNRAS.354.1093O}. Here, we discuss these effects for an idealized case of a plane-parallel shock in the ICM, having in mind the capabilities of X-ray calorimeters such as the \textit{XRISM} observatory \citep{2025PASJ...77S...1T} to complement various merger signatures with high-resolution spectroscopic information.

In the next section, we consider two models. 

The first model assumes that the downstream temperatures of all species are equal immediately after the shock, and only the ionization balance evolves with time. A parameter $\tau=n_e t\sim 10^{12}\,\rm cm^{-3}\,s$, where $n_e$ is the electron density, sets the characteristic timescale $t$ of non-equilibrium ionization (NEI) effects in this model, in particular, deviations of the relative line strengths from the Collisional Ionization Equilibrium (CIE) predictions for the electron temperature measured from the shape of the continuum spectrum. 

In the second model, the electrons and ions have different temperatures downstream of the shock and gradually equilibrate via Coulomb collisions. In this model,
\begin{itemize}
    \item electrons are heated adiabatically, while ions, on top of adiabatic heating, are heated in proportion to their mass. In addition to the NIE effects, line widths are now governed by the different ion temperatures. 
    \item ions equilibrate with each other first, lowering their temperatures to the temperature of protons on a similar timescale to the ionization equilibrium.
    \item electrons equilibrate with protons (and ions) on $\sim 10$ times longer timescale, 
    with the ionization equilibrium evolving along with the electron temperature, while the thermal broadening of ions follows the proton temperature.    
\end{itemize}

\begin{figure}
\centering
\includegraphics[angle=0,trim=1cm 5.5cm 1cm 2.5cm,width=0.9\columnwidth]{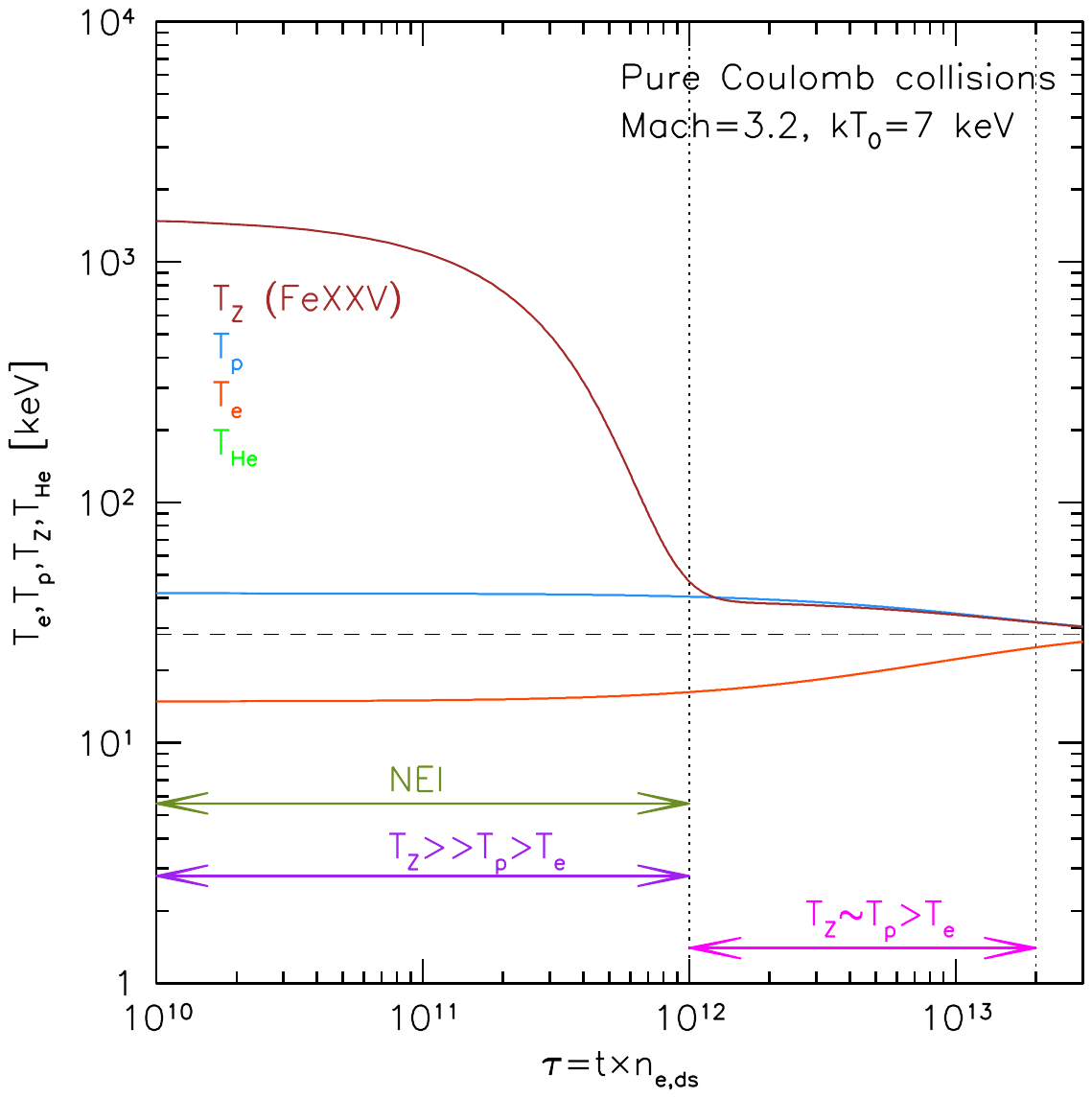}
\caption{Coulomb-collision-mediated temperature equilibration between electrons, protons, and He-like iron ions (Fe~XXV) in the limit of adiabatic heating of electrons at the shock.  The Fe~XXV ions' temperature $T_Z$ approaches $T_{\rm p}$ at $\tau\sim 10^{12}\,{\rm cm^{-3}s}$, and then evolves together with $T_{\rm p}$. The ``NEI'' timescale (see Fig.~\ref{f:ib_tmp} below) is approximately the same as the ``$T_Z\gg T_{\rm p}$'' timescale. 
Ions remain hotter than electrons up to $\tau\sim 2-3\times 10^{13}\,{\rm cm^{-3}s}$. Therefore, up to this value of $\tau$, the Fe~XXV ions are hotter than $T_{\rm e}$, and the thermal broadening of the Fe~XXV lines will be larger than the predictions based on the electron temperature $T_{\rm e}$ that can be derived from the shape of the continuum. The abundance of helium is set to zero for this plot (see Fig.~\ref{f:coulomb_he} for the impact of helium.}
\label{f:coulomb}
\end{figure}

\section{Model}
\label{s:model}
\subsection{Relevant physical processes}
As an illustrative example, we consider hot plasma with initial temperature $T_0=7\,{\rm keV}$. In the initial state, the electron and ion temperatures are equal ($T_e=T_i$), and the plasma is in the state of collisional ionization equilibrium (CIE) corresponding to this temperature\footnote{We use $T_i$ for the temperature of any ion, including protons. When needed, we use $T_{\rm p}$ for protons and $T_Z$ for ions of high-Z elements like iron.}. The plasma is optically thin, and the level population corresponds to the coronal approximation. A plane-parallel (collisionless) shock with $\mathscr{M}=3.2$ instantly changes the density and the mean plasma temperature according to the Rankine-Hugoniot conditions for the adiabatic index $\gamma=5/3$. For $\mathscr{M}=3.2$, the density and temperature jumps are $\rho_{\rm ds} \approx 3.09\rho_0$ and $T_{\rm ds} \approx 4.06 T_0$, with the subscript ``ds'' denoting downstream of the shock.

\subsection{Electron-ion temperature equilibration}

For the downstream temperatures of different species, we consider two cases:
\begin{itemize}
    \item A: $T_e=T_p=T_Z=T_{\rm ds}$ for all species.
    \item B: $T_e\neq T_p\neq T_Z$ and pure Coulomb collisions control the temperature equilibration. 
\end{itemize}
In the second case, we set the initial temperatures ``maximally different''. 
Namely, we assume that ion temperatures (on top of adiabatic heating) scale with the ion masses $T_i\propto m_i$, while the initial electron temperature corresponds to the adiabatic compression set by the density jump, i.e., $T_{e,\rm ds}=T_0\left ( \rho_{\rm ds}/\rho_0\right)^{(\gamma-1)}  \approx 2.12 T_0$ (for the compression ratio $3.09$). This assumption corresponds to the limit where each particle species thermalizes a fraction of the shock speed independently, without any exchange of energy among particle species.  It may occur in collisionless shocks, where the shock transition is mediated by plasma turbulence and electromagnetic fields instead of collisions between particles \citep[see, e.g.][and references therein]{2007ApJ...654L..69G,2008SSRv..134..141B,2015A&A...579A..13V,raymond17}.  With this definition, electrons, protons, and Fe~XXV ions will have the following temperatures immediately downstream: $T_{e,\rm ds}\approx 15\,\rm keV$,  $T_{p,\rm ds}\approx 43\,\rm keV$, and 
$T_{\rm Fe,\rm ds}\approx 1.5\,\rm MeV$, respectively. Unlike high-Mach-number shocks in young supernova remnants \citep[][]{2013SSRv..178..633G,2023ApJ...949...50R}, where the $T_{e,\rm ds}/T_{p,\rm ds}$ ratio can be very low (under the same assumptions), in clusters this ratio is modest by virtue of the limited sonic Mach number, so that the adiabatically heated electrons are only a factor of $\lesssim3$ colder than protons. The adiabatic heating of electrons at ICM shocks is also supported by observations \cite[e.g.,][]{2022MNRAS.514.1477R,2024ApJ...962..161S,2025ApJ...992...62N}.

Subsequent evolution of temperatures in case $B$ can be reasonably accurately described by the coupling of electrons and protons, and the ions and protons, namely
\begin{eqnarray}
    \frac{\mathrm{d}T_e}{\mathrm{d}t}=\nu_{\rm ep}n_p\left ( T_p-T_e\right )   
        \label{e:te_evol1}
    \\
     \frac{\mathrm{d}T_p}{\mathrm{d}t}=\nu_{\rm ep}n_e\left ( T_e-T_p\right )   
     \label{e:te_evol2}
     \\
     \frac{\mathrm{d}T_Z}{\mathrm{d}t}=\nu_{Ze}n_e\left ( T_e-T_Z\right )+  \nu_{Zp}n_p\left ( T_p-T_Z\right ) 
     \label{e:te_evol3}
\end{eqnarray}
where the $n_{\rm e}$, $n_{\rm p}$, and $n_Z$ are the number densities of electrons, protons, and heavy ions, respectively; all $\nu_{ab}$ terms define the rate of energy exchange between two species, $a$ and $b$ \citep{1962pfig.book.....S}. 
A more accurate treatment could involve helium and coupling between all species. This is done in the appendix~\ref{s:he}. However, for our illustrative example, the above three equations are fully sufficient. The first two describe the energy exchange between electrons and protons. Heavy ions are not very important in the overall energy budget due to their low abundance. Therefore, we can trace their temperature evolution due to collisions with electrons and protons (the 3rd equation), but we can ignore their back reaction on other species. Since we ignore the impact of He, it would be logical to ignore the first term in the r.h.s. of the 3rd equation, which is subdominant compared to the rate of energy exchange between iron and helium ions. We keep it to have a self-consistent approximation, where all ions heavier than a proton (by assumption) make a negligible contribution to the total energy density. As we illustrate in appendix~\ref{s:he}, this is sufficient for typical applications, unless very accurate predictions are needed.

The evolution of $T_{\rm e}$, $T_{\rm p}$, and $T_Z$ in our illustrative model is shown in Fig.~\ref{f:coulomb}. As expected, for pure Coulomb collisions, $T_Z$ approaches $T_{\rm p}$ on a timescale $t_{Z,p}$, which is an order of magnitude shorter than that of the electrons and ions equilibration. The corresponding value of $\tau_{Z,p}=t_{Z,p} n_e\approx t_{Z,p} n_p\lesssim 10^{12} \,{\rm cm^{-3}s}$, while  $\tau_{e,p}\sim 10^{13} \,{\rm cm^{-3}s}$. As we see below (Sect.~\ref{s:nei}), the typical value of $\tau_{\rm CIE}$ needed to establish ionization equilibrium is of the same order as $\tau_{Z,p}$ and smaller than $\tau_{e,p}$.
This means that for $\tau\gtrsim 10^{12}\,{\rm cm^{-3}s}$, the ion fractions will be close to the CIE and will follow the $T_e(t)$ evolution. The observed spectra (for these values of $\tau$) can, therefore, be represented as  
a superposition of CIE spectra with different plasma temperatures. 
However, the thermal line width of ions is set by $T_{Z}$, which can be larger than $T_{e}$ up to $\tau\sim 10^{13}\,{\rm cm^{-3}s}$. Therefore, the measured line width might be erroneously attributed to turbulence if only the iron lines are seen (one needs at least two elements with different masses to differentiate between turbulence and thermal broadening). We return to this issue in Sect.~\ref{s:diagnostic}.

We note that Coulomb collisionality is unlikely to govern all equilibration processes in the ICM.
Magnetic fields can effectively increase the rate of ion-electron collisions via scattering from plasma instabilities \citep{schekochihin2006} or anisotropic transport along field lines \citep{kunz2011}.
A reduced effective viscosity below the Coulomb level has been detected in the ICM via the suppression of Kelvin-Helmholtz instabilities in cold fronts \citep[e.g.,][]{2007PhR...443....1M,zuhone2016} and via the extension of density fluctuations power spectra into the viscous regime \citep{zhuravleva2019,heinrich2024}.
Thus, the actual isotropisation rate is likely to lie somewhere between cases A and B. This, however, does not necessarily mean that the temperature equilibration rate is also enhanced since the latter requires a transfer of energy from heavier to lighter particles (from ions to electrons) rather than the randomization of particles' velocities and the momentum exchange.

\subsection{NEI and He-like triplet}
\label{s:nei}
To study the time-varying ionization balance and line intensities, we make use of two collisional-radiative models, namely, 
the \texttt{CHIANTI} database \citep{2007A&A...466..771D,2021ApJ...909...38D} and the \texttt{NOMAD} code (\cite{Ralchenko_2001}, see Appendix for more details). As will be seen in the following, good agreement between two independent sets of simulations provides extra confidence in the derived results. 

\begin{figure}
\centering
\includegraphics[angle=0,trim=1cm 5.5cm 1cm 2.5cm,width=0.9\columnwidth]{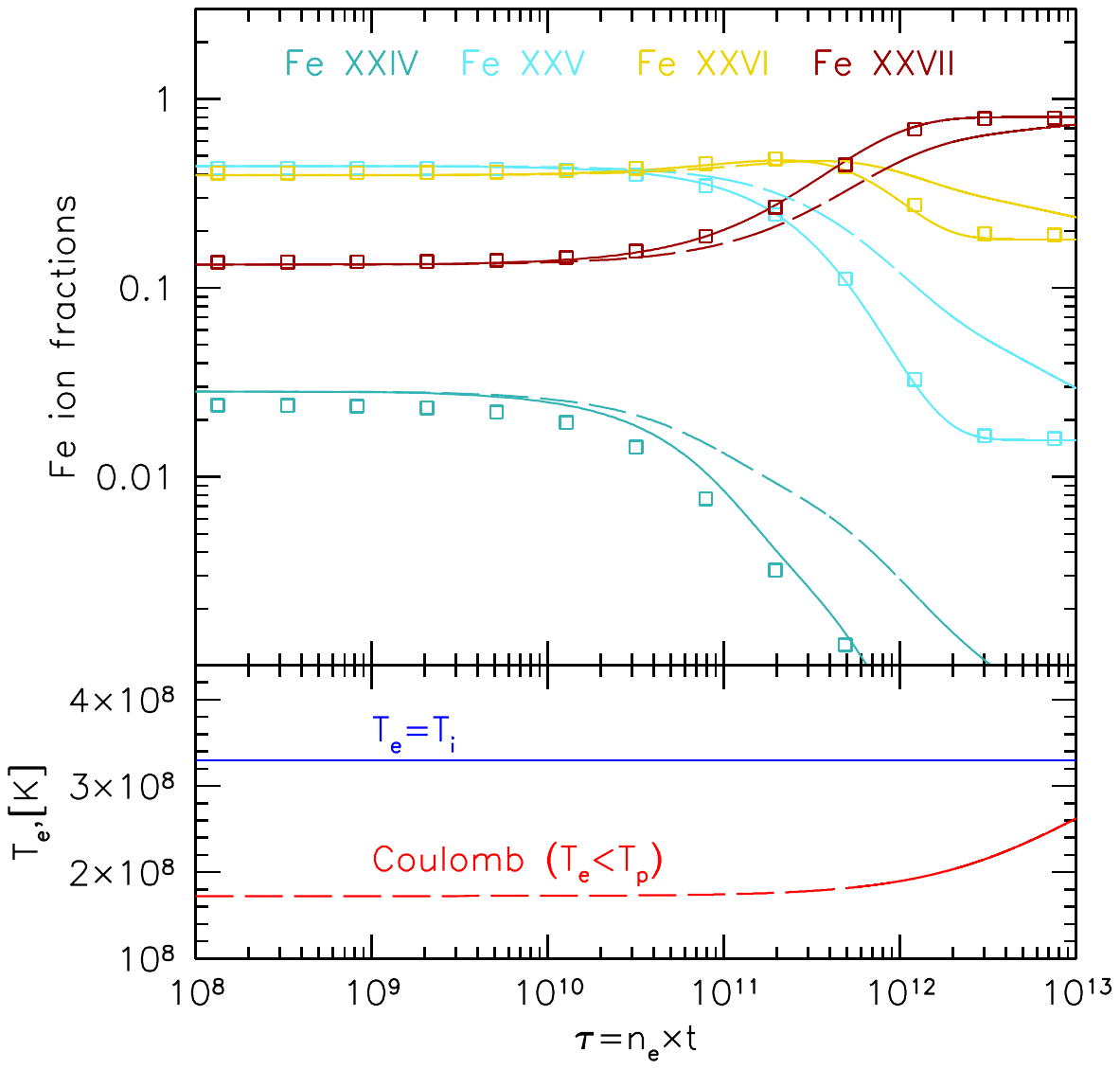}
\caption{Ionization balance and temperature evolution in the shocked medium. The bottom panel shows the $T_{\rm e}$ evolution: the blue solid line corresponds to the $T_e=T_i={\rm const}$ case, the red dashed line shows the ``Coulomb'' temperature equilibration starting from $T_e$ set by adiabatic compression in the $\mathscr{M}=3.2$ shock. In this case, $T_{\rm e}$ rises slowly and does not reach $T_{\rm p}$ even for $\tau=10^{12}\,{\rm cm^{-3}s}$. The upper panel shows the ionization balance evolution: solid lines illustrate the $T_e=T_i$ case, based on \texttt{CHIANTI} rates, open squares: \texttt{NOMAD}. Dashed lines show the ion fraction evolution when the electron temperature itself evolves with time, as shown in the lower panel. 
}
\label{f:ib_tmp}
\end{figure}

The time evolution of the iron ion fractions (Fe~XXIV--XXVII) is shown in Fig.~\ref{f:ib_tmp} (top) with the colored lines and symbols. The solid lines (\texttt{CHIANTI}) and open squares (\texttt{NOMAD}) correspond to the case when $T_e=T_i$, while the dashed lines illustrate the case when the electron temperature evolves with time due to Coulomb collisions. 

The impact of the non-equilibrium ionization on the line ratios was long recognized \citep[e.g.,][]{1978A&A....65..115M,1999LNP...520..189L}, in particular, for conditions relevant for Solar flares and laboratory plasma. In their calculations, the initial temperatures were typically low, and the role of Li-like ions, once the temperature increases suddenly, was particularly important for driving the $\mathtt{Z/W}$ flux ratio above the characteristic CIE values. In contrast, in massive, high-temperature clusters, the fraction of Li-like iron ions is lower in the initial state, and the increased rate of inner-shell ionizations of these ions (contributing to $\mathtt{Z}$  line) remains subdominant to the increased rate of Fe~XXV collisional excitations populating the $\mathtt{W}$ line. For convenience, the key properties of the Fe~XXV ``triplet'' (and Fe~XXVI Ly$\alpha$ doublet) lines are given in Table~\ref{t:lines}.

\begin{figure}
\centering
\includegraphics[angle=0,width=1.\columnwidth,trim=1cm 1.3cm 3cm 2cm,]{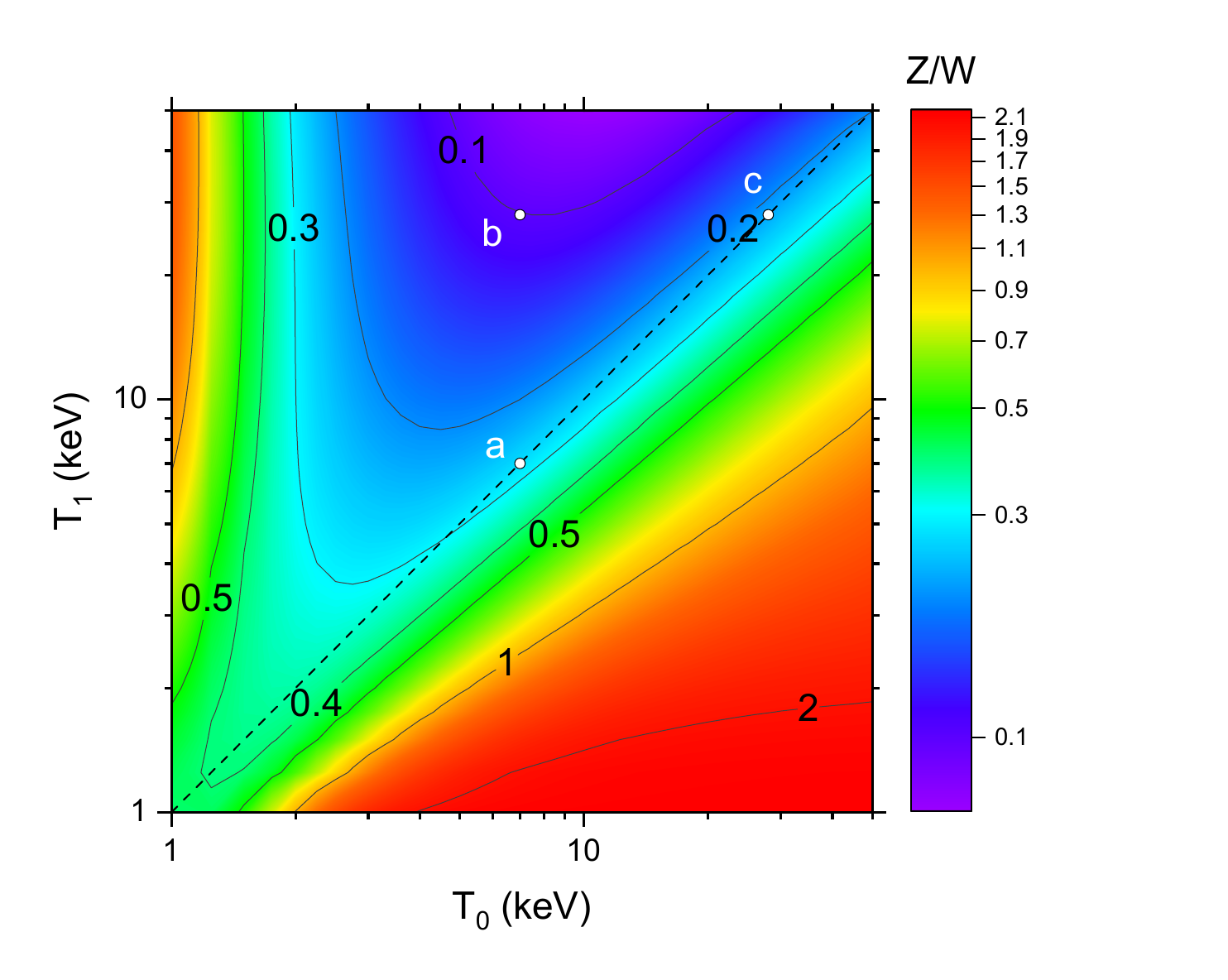}
\caption{$\mathtt{Z/W}$ line intensity ratio (colors, \texttt{NOMAD} simulation) for the instantaneous electron temperature change from $T_0$ to $T_1$ in the limit of small $\tau$. Locations $a$ and $c$ correspond to the flux ratios for the initial and final states with $T=7$ {\rm keV} and $T=28\,{\rm keV}$, respectively. The dashed diagonal line shows the locus of other equilibrium states with $T_1=T_0$. The location $b$ shows the intermediate case with  $T_0=7$ {\rm keV} and $T_1=28\,{\rm keV}$ and $\tau=10^{10}\,{\rm cm^{-3}s}$. The similar ratio is expected for a wide range of $\tau\lesssim 3\times 10^{10}\,{\rm cm^{-3}s}$. This transient state with the anomalously low $\mathtt{Z/W}$ ratio might be used as a signature of the ``freshly shocked'' hot NEI plasma.
}
\label{f:zw}
\end{figure}

The impact of the instantaneous change of electron temperature from $T_0$ to $T_1$ on the $\mathtt{Z/W}$ line intensity ratio is illustrated in Fig.~\ref{f:zw}. In this regime, the ionization balance corresponds to the CIE at temperature $T_0$, while the electrons' temperature is $T_1$. The upper-left corner corresponds to $T_1>T_0$ scenario, e.g., a shock, while the bottom right to the $T_1<T_0$ case, e.g., fast expansion. The former regime leads to a decrease in the $\mathtt{Z/W}$ ratio, while the latter pushes the ratio in the opposite direction. 

The evolution of the $\mathtt{Z/W}$ flux ratio as a function of the ionization parameter $\tau$ for our fiducial model is shown in Fig.\ref{f:lines}. The ratios were produced by \texttt{NOMAD} following the procedure described in the Appendix. As explained therein, during the NEI ``frozen ionization balance phase'' at $\tau \lesssim 10^{10}$ s/cm$^3$ the $\mathtt{Z/W}$ flux ratio is governed by several factors such as ion populations and modified rate coefficients for electron-impact processes. 

\begin{figure}
\centering
\includegraphics[angle=0,width=0.9\columnwidth]{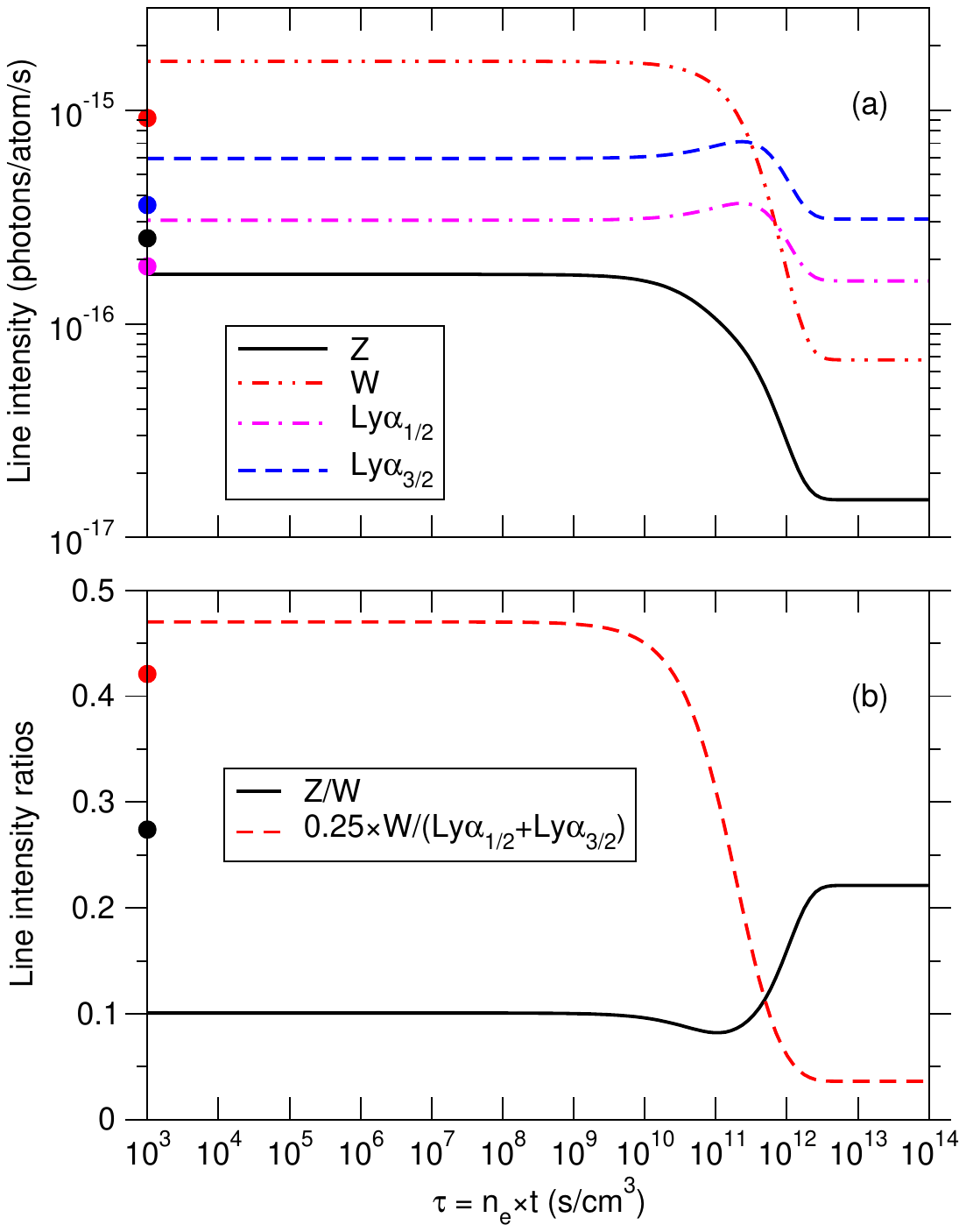}
\caption{\texttt{NOMAD} time-dependent (a) line intensities and (b) line intensity ratios for He- and H-like Fe ($T_0$ = 7 keV, $T_1$ = 28 keV, $T_e=T_{\rm p}$ case). The 0.25 scaling factor applied to the W/Ly$\alpha$ ratio is used to place all data on the same scale.} Solid circles present the corresponding values at $\tau$=0, i.e., $T_0$ = 7 keV.
\label{f:lines}
\end{figure}

\subsection{He-like vs H-like}
\begin{figure*}
\sidecaption
\includegraphics[angle=0,clip,trim=0.5cm 5.cm 2.5cm 2.5cm,width=6.0cm]{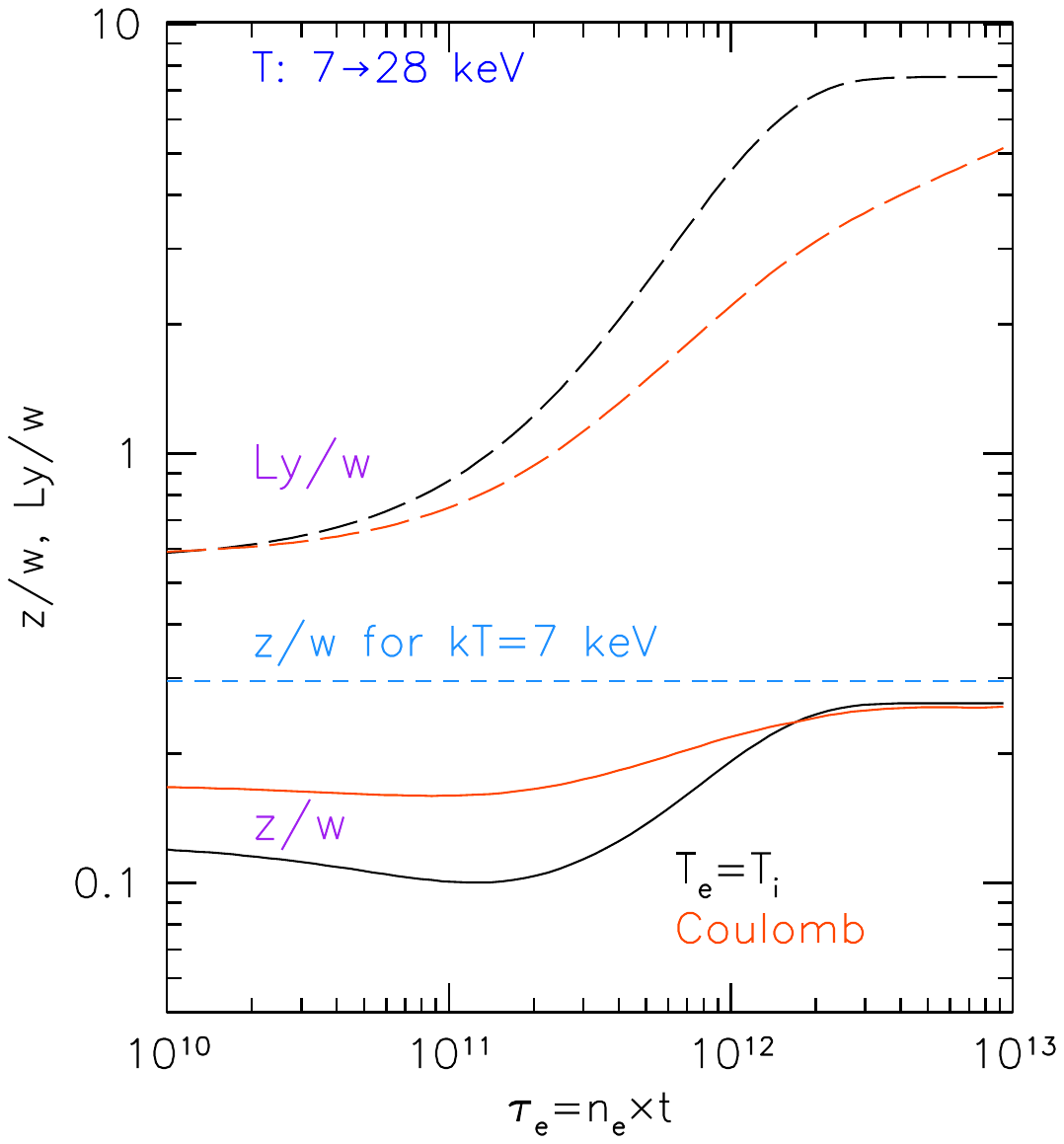}
\includegraphics[angle=0,clip,trim=0.5cm 5.cm 2.5cm 2.5cm,width=6.0cm]{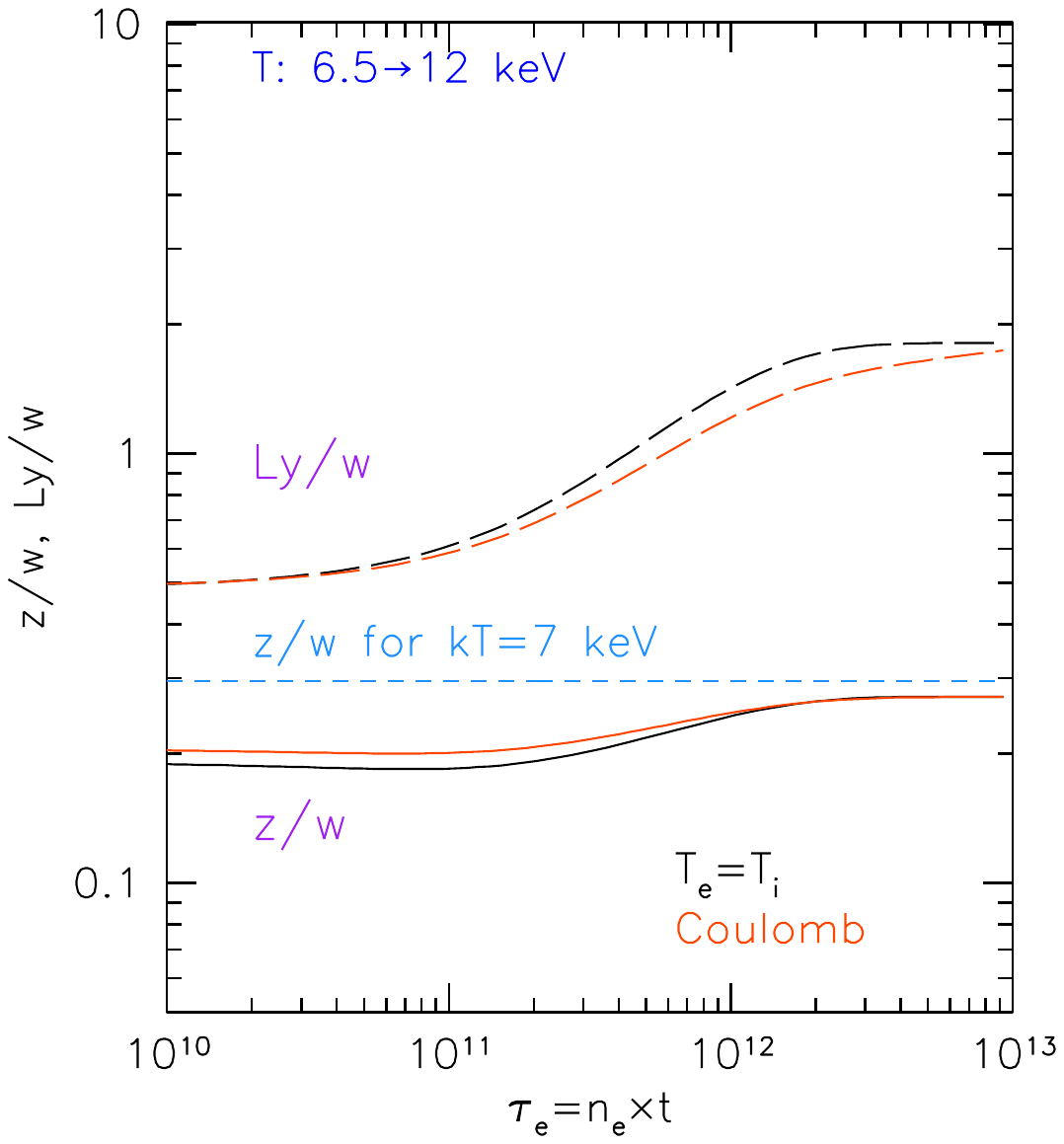}
\caption{{\bf Left:} $\mathtt{Z/W}$ (solid lines) and $\rm Ly/\mathtt{W}$ (dashed lines) ratios for the instantaneous (black) and Coulomb-collisions-mediated (red) electron temperature evolution from 7 to 28 keV (corresponding to $\mathscr{M}=3.2$ shock). The blue dashed line shows the ratio for $kT=7\,{\rm keV}$ CIE plasma. 
{\bf Right:} The same for the initial temperature of 6.5 keV and the final temperature of 12 keV (corresponding to $\mathscr{M}=1.8$ shock).
While the $\mathtt{Z/W}$ flux ratio differs for the $T_e=T_{\rm p}$ and $T_e<T_{\rm p}$ cases, it is always lower than the CIE-predicted values. The evolution of the $\rm Ly/\mathtt{W}$ is mostly driven by the time-dependent changes in the ionization balance.
\vspace{1cm}
}
\label{f:coulombww}
\end{figure*}

Another useful (and currently accessible) diagnostic is the ratio of the Fe~XXVI $\rm Ly_{\alpha}$ and the Fe~XXV triplet fluxes. For the CIE conditions, this ratio serves as a line-based temperature proxy. 
The NEI aspects of this flux ratio have also been considered  \citep[e.g.,][]{2010A&A...509A..29P,inoue2016}. For our baseline model with a rather hot ICM, the Boltzmann factor ($e^{-\Delta E/kT}$) is already close to unity ($\Delta E\ll kT$), and the flux ratio is mostly driven by the changes in the ionization fractions of iron, namely from the He-dominated state at the beginning to the H-dominated state at large $\tau$. This is illustrated with dashed lines in Fig.~\ref{f:lines}. A comparison of the line ratios for the $T_e=T_i$ and evolving $T_e$ cases is shown in Fig.~\ref{f:coulombww}. There, we also illustrate the impact of the shock strength (Mach numbers of $3.2$ and $1.8$) on the magnitude of the effect. Qualitatively, the suppression of the $\mathtt{Z/W}$ is always present, although a factor of $3$ decrease requires $\mathscr{M}\gtrsim 3$ shock.

The key role of NEI effects on the $\rm Ly_{\alpha}/\mathtt{W}$ ratio is that, due to the time-dependent changes in the ionization fractions, this ratio can have different values for the same electron temperature, derived from the shape of the continuum. This provides additional flexibility in describing spectra in merging clusters, and by itself serves as evidence for the recently shocked gas in the system.

We note in passing that none of the non-stationary effects considered in this study directly affects the $\rm Ly_\alpha$ doublet components ratio in Fe~XXVI. In the \texttt{NOMAD} calculations, this ratio ($\rm Ly{\alpha}_{3/2}/Ly{\alpha}_{1/2}$) is 1.94 (or 1.8 if a contribution of 1s-2s~M1 transitions is taken into account) and is about the same upstream and downstream of the shock. Therefore, the anomalies reported in \citet{2025ApJ...985L..20X} remain unexplained. Finite (and different) optical depths of the two components can modify the flux ratio, suppressing more the $\rm Ly{\alpha}_{3/2}$ if the scattering ``screen'' is between the bright core and the observer \citep[see examples in][]{2010SSRv..157..193C}. 
In particular, a configuration with velocity offset between the emitting and scattering materials by $\sim900$~km~s$^{-1}$ might result in the red-shifted $\rm Ly{\alpha}_{3/2}$ line falling in resonance with the $\rm Ly{\alpha}_{1/2}$ transition, and being subject more to the scattering \citep[as in the case considered in][]{2012AstL...38..443K}.
However, the estimated Coma optical depth in these lines is small \citep[e.g.,][]{2002MNRAS.333..191S} and does not resolve the anomaly \citep{2025ApJ...985L..20X}. We also note that, given sufficiently high photon count statistics, one can use the ratio of the Fe~XXV K$\beta$ to Fe~XXV K$\alpha-\texttt{W}$ line to estimate possible suppression of the resonant component (since this ratio is virtually insensitive to the ionization state and electron temperature of the plasma when $kT_e\gg E_{K\beta}-E_{W} \approx1$ keV).

\subsection{Line broadening}
\label{s:broadening}

The broadening of the Fe~XXV and Fe~XXVI lines is the direct proxy of ion motions. It can be subdivided into thermal motions of individual ions and macroscopic motions of the plasma. Since the velocity of thermal motions scales as $\left ( \frac{T_Z}{m_Z}\right )^{1/2}$, where $m_Z\approx 2Z m_{\rm p}$ is the mass of the ion, the contribution of thermal motions to the line width is the smallest for heavy ions, which, therefore, provide the best constraints on the ICM velocities. Most often, the CIE conditions are assumed and $T_Z=T_{\rm e}$. This allows constraining the temperature from the broad-band X-ray spectrum and the line ratios, predicting pure thermal broadening and determining an additional velocity broadening needed to describe the observed line width. These assumptions, however, may not be valid downstream of the shock. As shown in Fig.~\ref{f:coulomb}, the temperature of iron ions can be higher than the electron temperature for $\tau\lesssim 10^{13}\,{\rm cm^{-3}s}$, while for $\tau\lesssim 10^{11}\,{\rm cm^{-3}s}$, even the CIE assumption is not valid. Figure~\ref{f:broadening} illustrates these cases for our fiducial model with pure Coulomb collisions and no plasma bulk motions. In reality, supersonic bulk motions, contributing to the observed line width, must be present in the shock scenario unless the normal to the shock front is exactly in the sky plane. Essentially, the excess broadening of lines (e.g., in the $T_Z\propto m_ZT_{\rm p}$ scenario) is the result of isotropisation of the ions' velocity jump at the shock front. Therefore, the correct interpretation of the observed broadened lines requires some knowledge of the flow geometry.

\section{Discussion}
\label{s:discussion}

\subsection{Non-equilibrium vs equilibrium emission measure}
\label{s:neq_vs_eq_em}

\begin{figure}
\includegraphics[angle=0,trim=1cm 5.5cm 1cm 2.5cm,width=0.9\columnwidth]{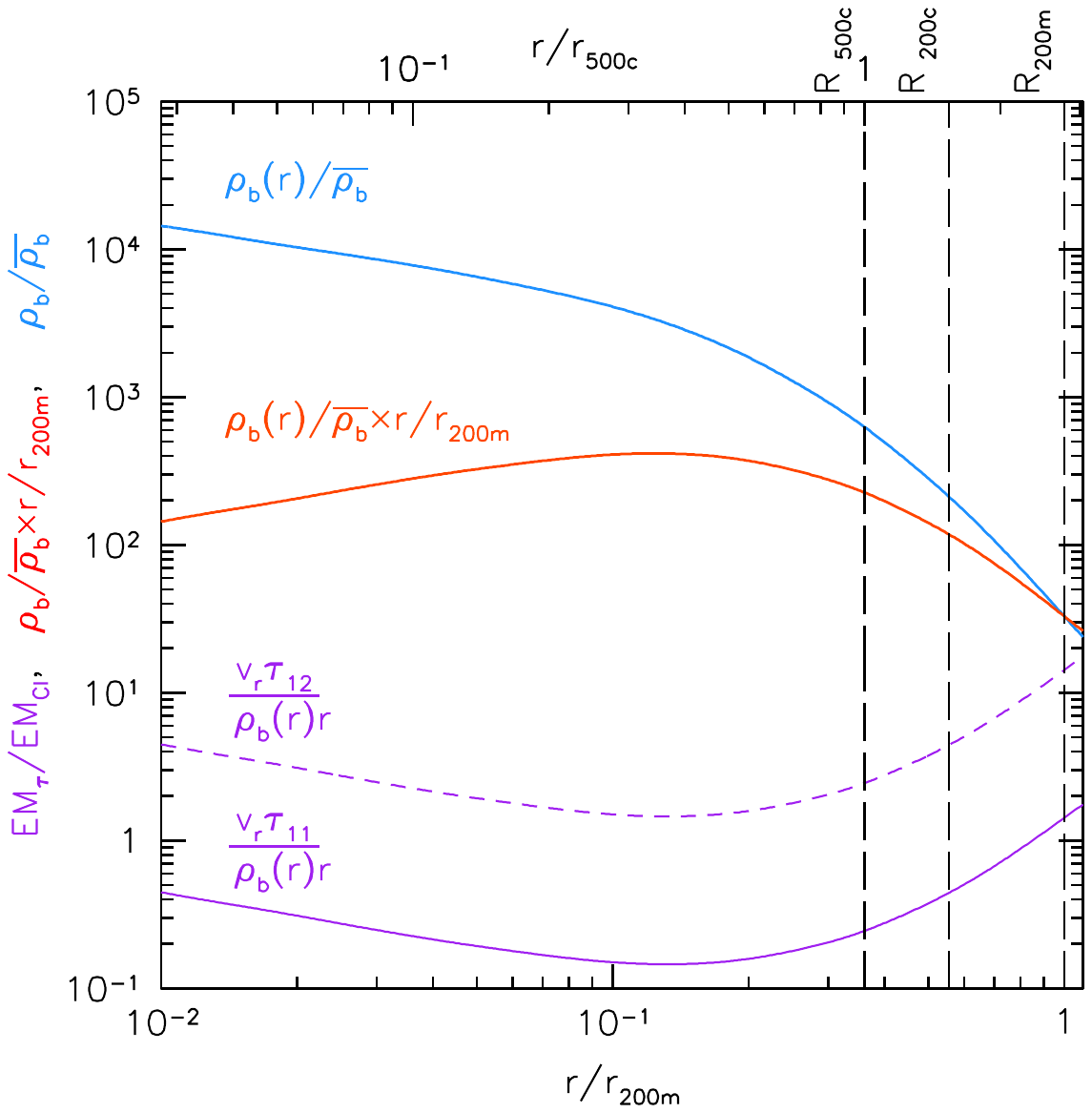}
\caption{Emission measure ${\rm EM}_\tau$ of the shocked gas with the ionization parameter $\tau$ smaller than a certain value relative to the emission measure ${\rm EM}_{\rm cl}$ of the unshocked gas at the same distance from the cluster center (see Eq.~\ref{e:em_ratio}). The purple solid and dashed lines show the ratio ${\rm EM}_\tau/{\rm EM}_{\rm cl}$ for $\tau=10^{11}$ and $10^{12}\,{\rm cm^{-3}s}$, respectively. For a shock velocity, the velocity $\varv_r$ of a point mass on a parabolic orbit was used. The blue and red lines show the   
overdensity of baryons as a function of radius and the product of the overdensity and the radius, respectively. The reciprocal of the latter quantity is the key quantity that determines the ratio ${\rm EM}_\tau/{\rm EM}_{\rm cl}$. From these plots, it follows that regions with $\tau\sim 10^{11} \,{\rm cm^{-3}s}$ can contribute $\sim 10\%$ to the line-of-sight emission measure, while for  $\tau\sim 10^{12} \,{\rm cm^{-3}s}$  the contributions of the shock and unshocked regions can be comparable. This conclusion holds essentially for any projected distance from the cluster center, although projection effects should also play a role.}
\label{f:emrat}
\end{figure}

We first make an order-of-magnitude comparison of the shocked gas contribution to X-ray spectra with that of the unshocked gas. Specifically, we are interested in regions having a characteristic ionization parameter $\tau$ comparable either to the NEI scale or to the $T_e\neq T_{\rm p}$ scale.
Consider a plane shock viewed along the normal to the shock front. The downstream density and velocity are $\rho_{\rm ds}=C\rho_0$ and $\varv_{\rm ds}=\varv_{\rm s}/C$, where $C$ is the compression factor. The length $l_{\tau}$ of the downstream region with the ionization parameter smaller than $\tau$ is:
\begin{eqnarray} l_{\tau}=\frac{\tau}{\rho_{\rm ds}}\varv_{\rm ds}=\frac{\tau}{\rho_{0}C}\frac{\varv_{\rm s}}{C}.
\end{eqnarray}
Therefore, the line-of-sight-integrated emission measure of this region is
\begin{eqnarray} 
{\rm EM}_{\tau}=l_\tau \left (\rho_0 C \right)^2= \varv_{\rm s} \tau \rho_{0}.
\end{eqnarray}
This value can be compared with the cluster emission measure along the line of sight
\begin{eqnarray}
   {\rm EM}_{\rm cl}\sim  \rho^2_{0} R_{\rm cl},
\end{eqnarray}
where $R_{\rm cl}$ is the characteristic size of the region dominating the X-ray emission.   
Thus, the ratio
\begin{eqnarray}
\frac{{\rm EM}_{\tau}}{{\rm EM}_{\rm cl}}=\frac{\varv_{\rm s} \tau }{\rho_{0} R_{\rm cl}}
\label{e:em_ratio}
\end{eqnarray}
determines the relative contributions of the non-equilibrium part of the shock-heated gas to the total cluster flux. 
For a typical cluster density profile, the distance from the center of the cluster $r$ can be used as $R_{\rm cl}$. Furthermore, the quantity $\rho_{0} R_{\rm cl}\sim \rho(r) r$ shows where in the cluster the impact of the shock is expected to be stronger. This is illustrated in Fig.~\ref{f:emrat}. There, we set $\varv_{\rm s}=\varv_{\rm r}$ - the velocity of a particle in a parabolic orbit in the NFW potential. We also set $\rho(r)$ to the mean density profile of simulated clusters from \cite{2021MNRAS.504.4649O}. Observationally, it is difficult to find examples of clusters in which the shock crosses the center now, since this phase is short-lived \citep[see, however,][for examples of merging clusters close to the pericenter passage]{2025A&A...693A..55L}. It is also difficult to get detailed spectra from the clusters' outskirts with the current generation of X-ray telescopes due to the very low gas density, hence low flux. So, the most promising are the intermediate regions within $R_{500c}$. As is clear from Fig.~\ref{f:emrat}, the contribution of the shock-heated gas at these radii is $\sim 10\%$ for $\tau\lesssim 10^{11}\,{\rm cm^{-3}s}$ and of the order of unity for $\tau\lesssim 10^{12}\,{\rm cm^{-3}s}$. Comparing with Fig.~\ref{f:coulomb}, we conclude that under favorable conditions, the contribution of the shock can be large enough to produce observable spectral signatures associated with the ionization parameter $\tau \sim 10^{11}\,{\rm cm^{-3}s}$.

\subsection{Diagnostic}
\label{s:diagnostic}
\begin{figure*}
\sidecaption
\includegraphics[angle=0,clip,trim=0.5cm 5.cm 1.5cm 2.5cm,width=6.cm]{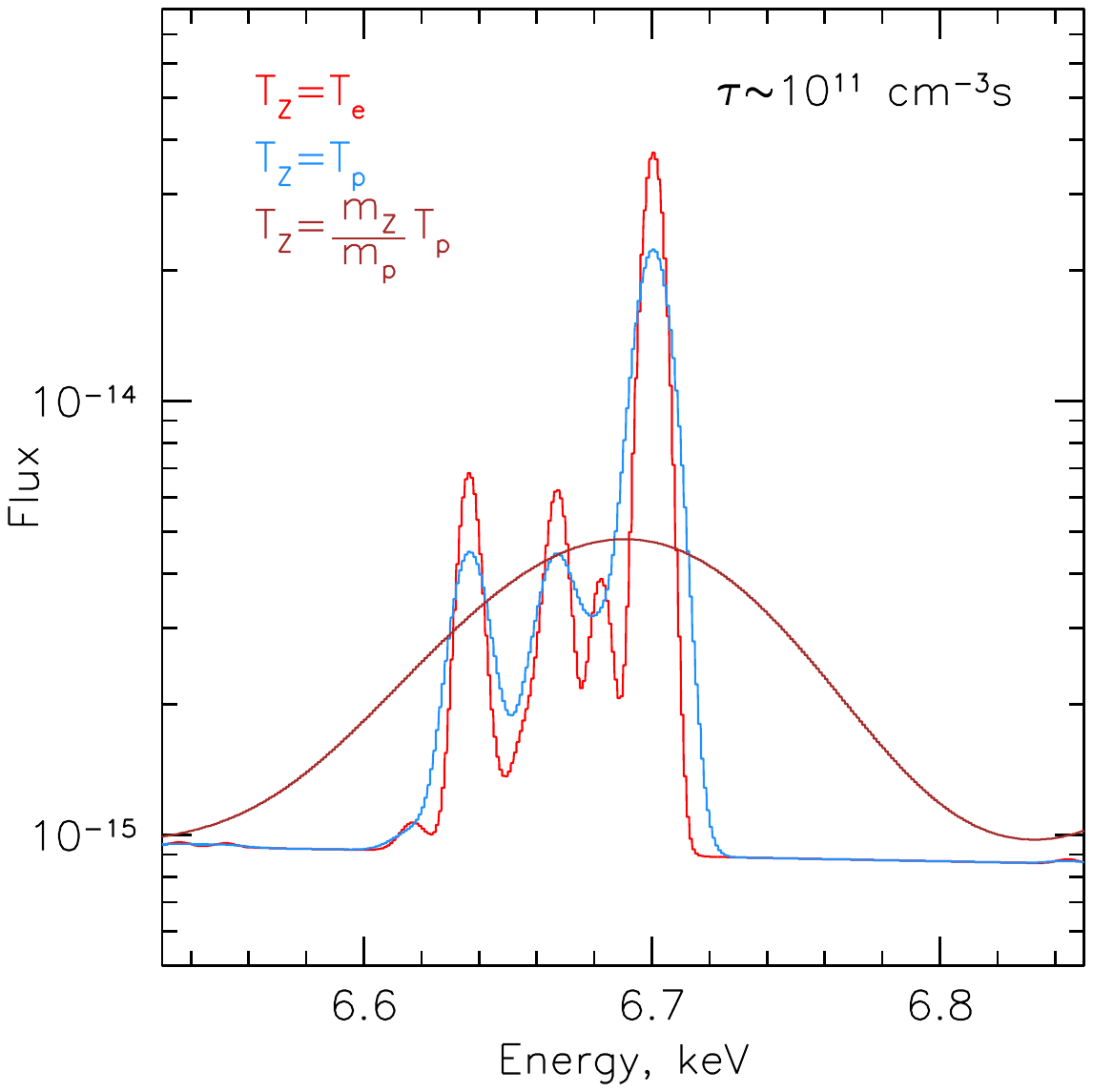}
\includegraphics[angle=0,clip,trim=0.5cm 5.cm 1.5cm 2.5cm,width=6.cm]{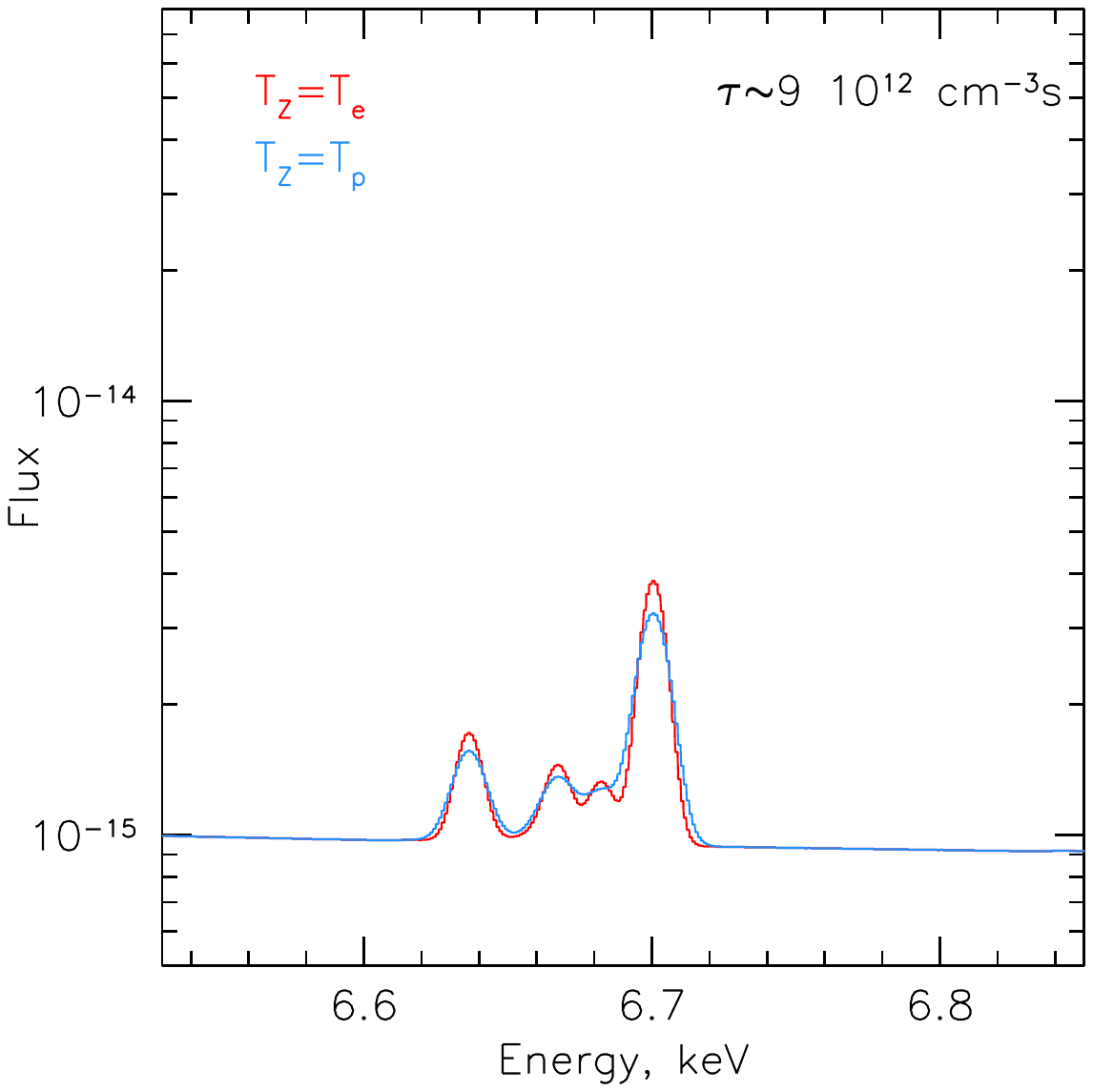}
\caption{Triplet spectrum for different assumptions on the ion temperature and different ionization parameter $\tau$. \textbf{Left:} 
$\tau=10^{11}\,{\rm cm^{-3}s}$. The red, blue, and brown curves show $T_Z=T_{\rm e}$, $T_Z=T_{\rm p}$, and $T_Z=\frac{m_Z}{m_p}\approx 2Z\times T_{\rm p}$ in a shocked plasma with pure Coulomb energy exchange between species. \textbf{Right:} $\tau=9\times 10^{12}\,{\rm cm^{-3}s}$. For this value of $\tau$, $T_Z=T_{\rm p}$, but it is still larger than $T_{\rm e}$, leading to marginally broader lines. For these large values of $\tau$, the ionization fractions are tracing the time evolution of $T_e$, while ion temperature tracks $T_p$. As a result, the fluxes of the triplet ($T_{\rm Z}=T_{\rm p}$ case blue line) can be reproduced in XSPEC with the \texttt{BAPEC} model with $kT=22\,\rm keV$ (instead of 28~keV) and an extra gaussian velocity broadening of $\sigma=180\,\rm km\,s^{-1}$. The other model ($T_{\rm Z}=T_{\rm e}$; red line) is reproduced by the \texttt{BAPEC} model with $kT=22\,\rm keV$ and no velocity broadening.}
\label{f:broadening}
\end{figure*}

The baseline spectral model used to describe emission from galaxy clusters is the optically thin thermal plasma emission model in collisional ionization equilibrium. The second often used model is a two-temperature CIE model that accounts for plasma with different temperatures observed along the same line of sight. The features associated with deviations from ionization or electron/ion temperature equilibria might not be adequately captured by the two-temperature model. As discussed in Sect.~\ref{s:model}, there are two characteristic values of $\tau$, which are associated with specific non-equilibrium features in the X-ray spectra.  Namely,
\begin{enumerate}
    \item For $\tau\lesssim 10^{12}\,{\rm cm^{-3}s}$, the $\mathtt{Z/W}$ ratio in the Fe~XXV triplet can be anomalously low due to NEI effects (see Fig.~\ref{f:zw}).
    \item For $\tau\lesssim 10^{12}\,{\rm cm^{-3}s}$, the line broadening (for Fe) can be extremely high, due to the higher temperature of massive ions in the absence of efficient ion/proton temperature equilibration (see the left panel of Fig.~\ref{f:broadening}).
    \item For $\tau\lesssim 10^{13}\,{\rm cm^{-3}s}$, the line broadening (for Fe) can be larger than the thermal broadening corresponding to the observed electron temperature in the regime, when $T_i=T_p>T_{\rm e}$, 
    in the absence of efficient electron/proton temperature equilibration (see the right panel of Fig.~\ref{f:broadening}).  
\end{enumerate}
The first item in the above list is associated with the ionization timescale of He-like ions and should be present even if $T_e=T_i=T_{\rm p}$ (see Fig.~\ref{f:coulomb}). The broadening (items 2 and 3), on the contrary, strongly depends on the initial temperatures of different species immediately downstream of the shock and on the efficiency of subsequent temperature equilibration. As discussed in Sect.~\ref{s:broadening}, the ability to differentiate the contributions of the ICM motions from the ions' thermal motions downstream of the shock requires explicit knowledge/model of the shock geometry.

Observationally, supernova remnant shocks show ion temperatures that are mass-proportional in high Mach number shocks, but near equilibration in lower Mach number shocks \citep{raymond17}.   Shocks in the solar wind show a great deal of scatter in the ratio of $T_i / T_{\rm p}$ \citep{korreck07}, but $T_i$ is generally greater than $T_{\rm p}$.  PIC simulations show close to mass-proportional thermal core temperatures for He and CNO in a  $\mathscr{M}=$40 shock \citep{caprioli25}.  Observations of shocks in supernova remnants indicate fairly complete electron-ion equilibration at low Mach numbers, but $T_{\rm e}$ only a few percent of $T_{\rm p}$ at high Mach numbers \citep{2012A&ARv..20...49V, 2013SSRv..178..633G, yamaguchi14, 2023ApJ...949...50R}.  Solar wind shocks show the same general trend, but with large scatter \citep{wilson20}, and $T_e/T_{\rm p}$ seems to rise slowly at high Mach numbers in PIC simulations as well as in solar wind shocks \citep{bohdan20, hanusch20, tsiolis21}.

\subsection{Radio-guided selection of potential sites of NEI}

Identifying regions that may exhibit departures from ionization equilibrium or differences between electron and ion temperatures benefits from accurate knowledge of the shock front location. However, detecting shocks in X-ray observations is often challenging due to low surface brightness and projection effects. Radio relics -- arc-like, polarized structures extending up to 1--2~Mpc in galaxy clusters -- are widely interpreted as tracers of shock fronts \citep[for a review see][]{2019SSRv..215...16V}. At these locations, non-thermal particles are (re)accelerated into a power-law energy distribution \citep[e.g.,][]{1983RPPh...46..973D,2017MNRAS.472.3605L,2020A&A...642L..13R} and subsequently advected downstream of the shock, where they cool on characteristic timescales given by
\begin{equation}
t_{\rm cool} [{\rm yr}] = 1.0\times10^{9}\,\frac{B_{\rm \mu G}^{1/2}}{B_{\rm \mu G}^2 + B_{\rm CMB,\mu G}^2} \left[(1+z)\nu_{\rm GHz}\right]^{-1/2},
\end{equation}
with the observed radio frequency $\nu_{\rm GHz}$ in GHz, magnetic filed strengths $B_{\rm \mu G}$ and $B_{\rm CMB,\mu G}$ in $\mu$Gauss, and $z$ being the redshift of the cluster. Here, $B_{\rm CMB}\approx 3.25\times (1+z)^2\rm \mu G$ is set by the condition that $B_{\rm CMB}^2/8\pi$ is equal to the Cosmic Microwave Background (CMB) energy density at redshift $z$.

This cooling leads to a progressive steepening of both the electron energy distribution and the observed radio spectrum over a characteristic length scale of $t_{\rm cool} \varv_{\rm ds}$. 
For typical ICM conditions, with magnetic field strengths of a few~$\mu$G and downstream velocities $\varv_{\rm ds} \sim 10^{3}$~km~s$^{-1}$, the radio emission at $\sim$1~GHz is expected to extend up to $\sim$50~kpc behind the shock front.

Regions with relatively flat spectral indices are therefore expected to lie close to the shock front, typically within a few tens of kpc. Selecting such regions may thus provide a practical way to isolate the immediate post-shock environment, particularly when the shock location cannot be robustly constrained from X-ray observations.

\subsection{Promising targets for search of NEI signatures}

At $z\sim0$, for the gas near $R_{500c}$, the value of $\tau\sim 10^{12}\,{\rm cm^{-3}s}$ corresponds to the timescale of $\sim 300\,{\rm Myr}$. It is short enough to have a relatively small impact on the properties of an average cluster. Therefore, favorable conditions are needed to make the discussed features observable, such as ongoing mergers of massive, i.e., high-temperature, clusters.

Here, we mention a few clusters, which might be good targets for the search for non-equilibrium features:
\begin{itemize}
    \item Bullet cluster (1E~0657$-$56)
    at $z = 0.296$ provides a canonical example of a strong  $\mathscr{M}\approx 3$ shock \citep{2002ApJ...567L..27M,2010arXiv1010.3660M}. A combination of X-ray and SZ data provides hints for $T_e<T_p$ \citep{2019A&A...628A.100D}.  
    \item El Gordo (ACT-CL~J0102-4915) at $z=0.87$ is another textbook example of merging galaxy clusters -- a rare, massive system at high redshift \citep{Menanteau2012}. Numerical modeling suggests that it is observed close to the plane of the sky, with a shock Mach number of $\sim4$, and approximately $100-200\,{\rm Myr}$ after the pericentric passage \citep{Zhang2015}.
    \item Peanut cluster (CL0238.3+2005) 
    at $z\approx 0.42$, $kT\approx 10\,{\rm keV}$ is a merger close to the pericenter passage with a line-of-sight difference between two subclusters of $\sim 2000 \, \rm km\,s^{-1}$ \citep{2025A&A...693A..55L,2026JHEAp..5300632Z}.
    \item Abell 2034 
    at $z\approx0.113$ is $kT\sim 8$ keV cluster undergoing a near plane-of-sky head-on merger. A prominent shock front lies $\sim 400$ kpc from the cluster center \citep{owers2013,monteiro-oliveira2018}. A2034 was recently observed by \textit{XRISM} \citep{2026arXiv260427161H}
    who found broader emission lines than have been inferred in other clusters so far. There are also indications that the $\mathtt{Z/W}$ ratio is low, possibly indicating an NEI state of the line-emitting plasma. However, this state is detected only at $\lesssim2\sigma$ significance in the current data.
    The He$\alpha$ and Ly$\alpha$ emission lines are consistent with a $M\approx1.8$ shock that heats the ICM from 6.5 keV to 12 keV. The evolution of the $\mathtt{Z/W}$ ratio for this shock is plotted in the right panel of Figure \ref{f:coulombww}. The apparent $\sim 470$ km/s velocity dispersion measured in this cluster could also be explained by a high Fe ion temperature of up to $T_Z\approx85$ keV.
    \item MACS\,J0717.5+3745 at $z=0.5458$, $kT\approx12$~keV with the suggested  line-of-sight velocity difference of $\sim 3000\, \rm km\,s^{-1}$ \citep{2017A&A...598A.115A}.
\end{itemize}

\section{Conclusions}

We discuss the spectral signatures of non-equilibrium ionization and the inequality of the electron and ion temperatures in the ICM, arising due to the propagation of merger shocks. We focus on the He-like iron triplet in hot plasma lacking Li-like ions, and the following features:

\begin{itemize}
    \item The reduced ratio of the forbidden to resonant line $\mathtt Z/W$ arising in plasma dominated by He- and H-like ions when the electrons' temperature increases suddenly. The duration of this phase is $t\sim 30 \left( \frac{n_e}{10^{-3}\,{\rm cm^{-3}}}\right)^{-1}{\rm Myr}$. For the $\mathscr{M}\sim 3$ shock, this translates into a physical size of the region downstream of the shock $\sim 50\,{\rm kpc}$ in a $kT\sim 10\,{\rm keV}$ cluster (Fig.~\ref{f:coulombww}).
    \item On a similar time scale, He-like lines will be boosted relative to H-like lines expected in CIE conditions for the values of $T_{\rm e}$ derived from the shape of the observed continuum. This happens due to the ``under-ionized'' state of plasma.
    \item Increased line broadening compared to expectations based on the $T_i=T_{\rm e}$ assumption. For the case when the Coulomb scattering is the sole mechanism that equilibrates the temperatures of different species, the lines can be very broad for  $t\sim 30 \left( \frac{n_e}{10^{-3}\,{\rm cm^{-3}}}\right)^{-1}{\rm Myr}$ when $T_i > T_p > T_{\rm e}$ and moderately broad for  $t\sim 300 \left( \frac{n_e}{10^{-3}\,{\rm cm^{-3}}}\right)^{-1}{\rm Myr}$, when $T_i = T_p > T_{\rm e}$. Ignoring this effect might lead to overestimation of the level of turbulence downstream of the shock (Fig.~\ref{f:broadening}).
     \item The ``magnitude'' of these transient features in the X-ray spectra scales linearly with density (rather than the density squared) because the duration of the transient phase is inversely proportional to density (Sect.~\ref{s:neq_vs_eq_em}, Fig.~\ref{f:emrat}).
\end{itemize}
These spectral signatures come on top of the line shift \& broadening and the structures in the X-ray surface brightness due to the shock-generated discontinuities in the gas velocity and density, respectively. With the advent of \textit{XRISM} Observatory \citep{2025PASJ...77S...1T}, it might be possible to identify spectral signatures specific to the shocks and place constraints on the electrons heating at the shock front and the temperature equilibration rates.

\begin{acknowledgements}
The work of Yu.R. was supported by NASA under award number 80GSFC24M0006.
IK was supported by the Simons Foundation via the Simons Investigator Award to A. A. Schekochihin.
CZ acknowledges the support of the Czech Science Foundation (GACR) Junior Star grant no. GM24-10599M. AH and IZ were partially supported by NASA grant number 80NSSC25K7693. 
\end{acknowledgements}

\bibliographystyle{aa}
\bibliography{ref} 
\begin{appendix}
\section{\texttt{NOMAD} simulations}
The calculations of ionization balances and line intensities were performed with the time-dependent collisional-radiative code \texttt{NOMAD} \citep{Ralchenko_2001}. The model includes ionization stages from Li-like Fe$^{23+}$ to bare nucleus with the total number of levels of about 1700. In particular, the singly excited levels with $n \leq 10$ and doubly excited autoionizing levels $1snln'l'$ and $nln'l'$ with $n \leq 3$ and $n' \leq 5$ were included for Li- and He-like ions. The atomic data (e.g., energy levels, radiative transition probabilities, collisional cross sections, autoionization probabilities, etc.) for simulations were calculated using Flexible Atomic Code \citep{gu_flexible_2008} based on the relativistic model potential method. 

\begin{figure}
\centering
\includegraphics[angle=0,width=0.9\columnwidth]{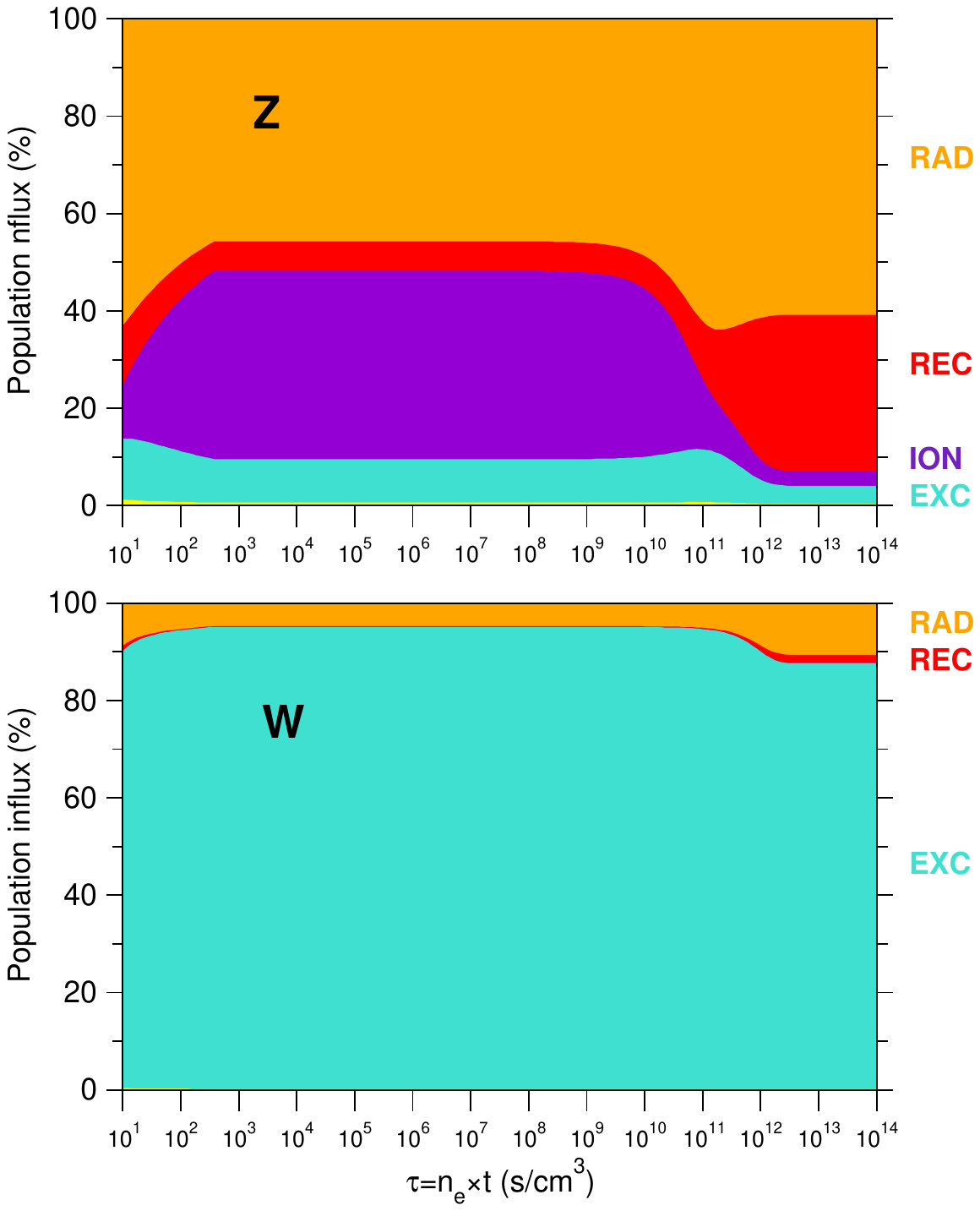}
\caption{Relative population influxes for $\mathtt{Z}$ and $\mathtt{W}$ lines ($T_0$ = 7 keV, $T_1$ = 28 keV, $T_e=T_{\rm p}$ case). RAD: radiative cascades, REC: direct radiative recombination from the H-like ground state, ION: inner shell ionization from the Li-like ion, EXC: electron-impact excitation.}
\label{f:fluxes}
\end{figure}

\begin{figure}
\centering
\includegraphics[angle=0,width=0.9\columnwidth]{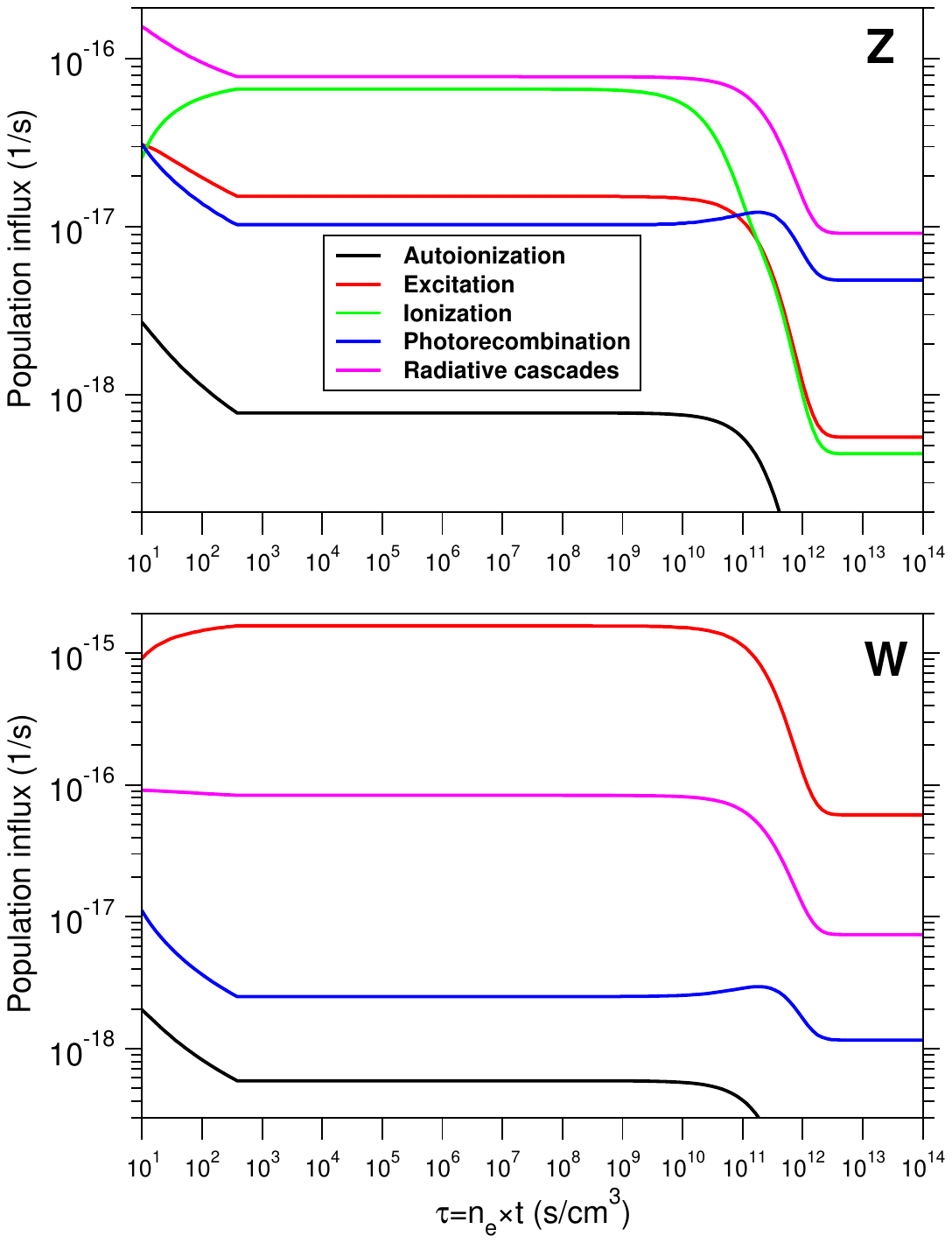}
\caption{Absolute population influxes for $\mathtt{Z}$ and $\mathtt{W}$ lines ($T_0$ = 7 keV, $T_1$~= 28 keV, $T_e=T_{\rm p}$ case).}
\label{f:fluxes_abs}
\end{figure}

In the first set of calculations, the evolution of the system of levels was determined from the rate equation:

\begin{equation}
    \frac{d\hat{N}(t)}{dt}=\hat{A}(t) \cdot \hat{N}(t)
\end{equation}
where $\hat{N}(t)$ is the vector of level populations over all ions ($\Sigma_i N_i(t)=1$) and $\hat{A}(t)$ is the rate matrix. The initial condition $\hat{N}(t=0)$ corresponds to the steady state equilibrium at $n_{\rm e}$ = 10$^{-3}$ cm$^{-3}$ and $T_{\rm e}$ = 7 keV. 
The time steps were chosen on a logarithmic scale via: $t_0$ = 0, $t_1$ = 10$^4$ s, and $t_{i+1}$ = $t_i \cdot 1.2$  for i=1..169 ($t_{169}$ = 2.408 $\cdot$ 10$^{17}$ s). At each time step between $i$=0 and  $i$=21, the electron temperature increased by 1 keV until reaching 28 keV at $t_{21}$ = 3.834 $\cdot$ 10$^5$ s, after which it was kept constant. The calculated ion populations and line intensities for He-like and H-like lines are shown in Figs. \ref{f:ib_tmp} (open squares) and \ref{f:lines}, respectively. The relative population influxes for $\mathtt{Z}$ and $\mathtt{W}$ lines are presented in Fig.~\ref{f:fluxes} while the absolute influxes are given in Fig.~\ref{f:fluxes_abs}. For the frozen ionization distribution at relatively small $\tau$, an instantaneous increase of the electron temperature results in a mildly enhanced excitation influx for the $\mathtt{W}$ line. Although $\mathtt{Z}$ also experiences an increased influx due to the inner-shell ionization from the ground state of 
Fe$^{23+}$, the relatively small population of the latter cannot provide an influx increase comparable to that for $W$. Hence, the $\mathtt{Z/W}$ ratio decreases as compared to the low-temperature condition.

For the second set of calculations, the initial temperature ($T_0$) was set in the 1..50 keV range. Then, the temperature ($T_1$) was changing from 1 keV to 50 keV with a time step of 10$^4$ s. This time step is too short to modify abundances of the ions involved while long enough to ensure quasi-steady-state equilibration between the excited states within an ion, which is established on the times compared to the longest lifetimes (cf. $\tau \simeq 5 \cdot 10^{-9}$~s~cm$^{-3}$ for the $Z$ line). Then, the corresponding $\mathtt{Z/W}$ ratio was determined for the combination of ($T_0$,$T_1$) as shown in Fig.~\ref{f:zw}. After the fast rise of electron temperature, the contribution of the inner shell ionization from the ground state $1s^22s$ of the Li-like ion to the $Z$ influx increases by about a factor of four and remains high until the progressive ionization begins to deplete Li-like ions at about $\tau \sim 10^{10}$~s~cm$^{-3}$ (Figs.~\ref{f:fluxes} and \ref{f:fluxes_abs}). 

\section{Reference list for the relevant lines}

In Table~\ref{t:lines}, we list key parameters of the lines considered in this study for reference.

\begin{table*}[]
\caption{Key parameters of the most relevant lines.}
    \centering
\begin{tabular}{lcccl}\hline\hline
\\
{Spectroscopic} &{Transition}&{Type}&{Energy, keV}&{Transition}\\
{symbol} &{}&{}&{}&{probability, s$^{-1}$}\\		
[2 mm]\hline \\
{Fe XXV $K_{\alpha}-\mathtt{W}$ }&{$1s^{2}$ $^1S_0$ - $1s2p$ $^{1}P_{1}$}&{E1}&{6.700}&{$4.57\times 10^{14}$}\\
{Fe XXV $K_{\alpha}-\mathtt{X}$}&{$1s^{2}$ $^1S_0$ - $1s2p$ $^{3}P_{2}$}&{M2}&{6.682}&{$6.64\times10^{9}$}\\
{Fe XXV $K_{\alpha}-\mathtt{Y}$}&{$1s^{2}$ $^1S_0$ - $1s2p$ $^{3}P_{1}$}&{E1}&{6.668}&{$4.42\times 10^{13}$}\\
{Fe XXV $K_{\alpha}-\mathtt{Z}$}&{$1s^{2}$ $^1S_0$ - $1s2s$ $^{3}S_{1}$}&{M1}&{6.634}&{$2.12\times 10^{8}$}\\[2mm]
{Fe XXVI $Ly{\alpha}_{3/2}$}&{$1s$ $^2S_{1/2}$ - $2p$ $^{2}P_{3/2}$}&{E1}&{6.973}&{$2.8370 \times 10 ^{14}$}\\
{Fe XXVI $Ly{\alpha}_{1/2}$}&{$1s$ $^2S_{1/2}$ - $2p$ $^{2}P_{1/2}$}&{E1}&{6.952}&{$2.8731\times10^{14}$}\\
\hline
\end{tabular}
    \tablefoot{All data (energies rounded to 1 eV) are from NIST Atomic Spectra Database \citep{NIST_ASD}. 
    }
    \label{t:lines}
\end{table*}

\section{Role of Helium in temperature equilibration}
\label{s:he}

\begin{figure}
\centering
\includegraphics[angle=0,trim=1cm 5.5cm 1cm 2.5cm,width=0.9\columnwidth]{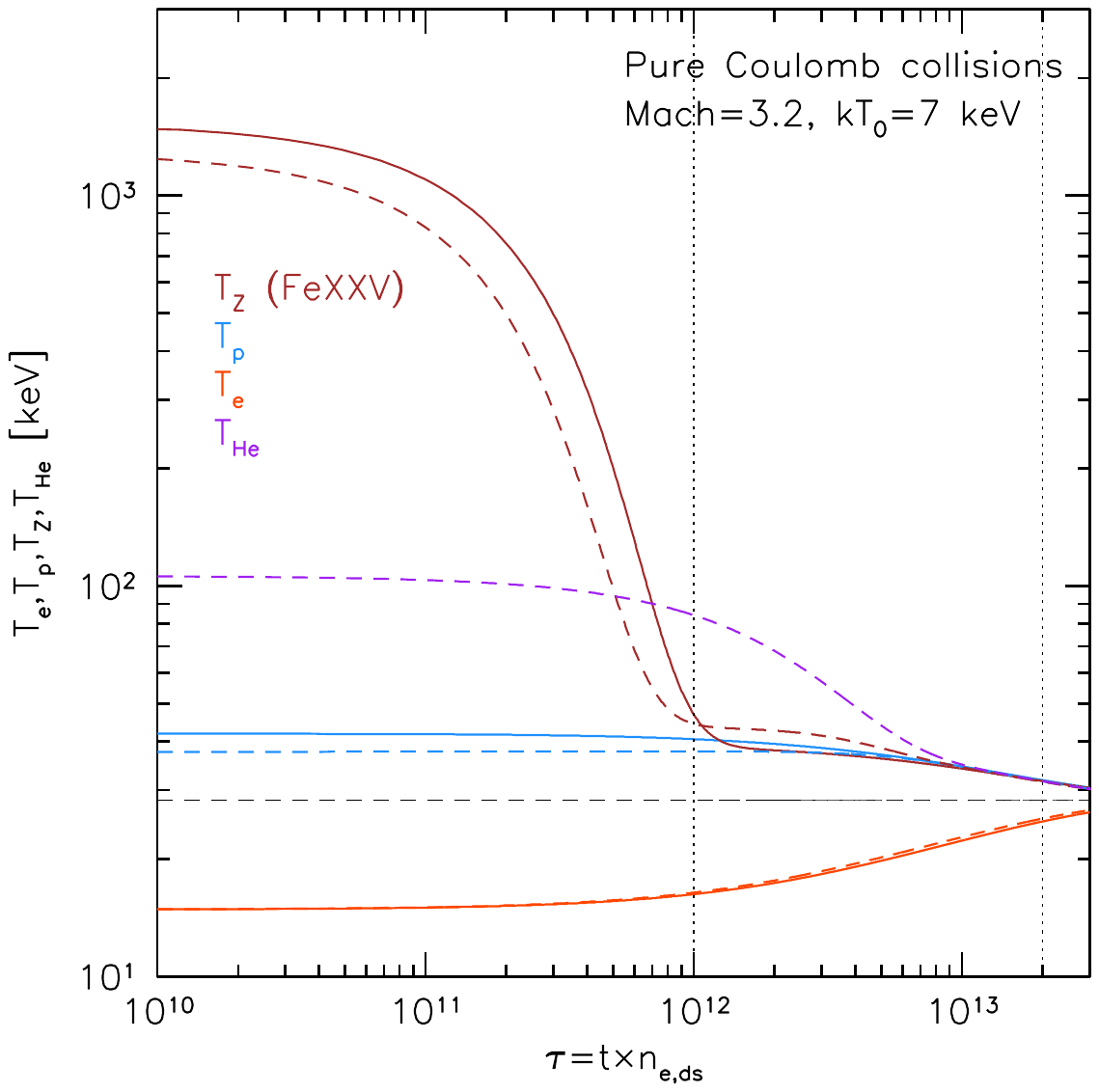}
\caption{Impact of helium on the temperature equilibration. The solid lines correspond to the case with no helium, while the dashed lines show the temperature evolution when the He abundance is set to $0.084$ of hydrogen \citep{2009LanB...4B..712L}.
The main effect of helium is (i) the increase of the mean atomic weight of plasma and, as a consequence, lower downstream temperature of ions  (visible for $\tau<10^{12}\,\rm cm^{-3}\,s$) and (ii) keeping Fe ions slightly hotter at $\tau>10^{12}\,\rm cm^{-3}\,s$, since the helium ions have the longest equilibration time.} 
\label{f:coulomb_he}
\end{figure}

In this section, we verify the impact of helium  
on the ion temperature evolution, considering energy exchange between electrons, protons, helium ions, and ions of Fe~XXV. The same assumption of a pure adiabatic heating of electrons at the shock is made. The subsequent evolution is mediated by Coulomb collisions.
Fig.~\ref{f:coulomb_he} compares the evolution with helium (dashed lines) and without helium (solid lines). 
In the former case, the initial downstream temperature of ions is lower because of the increased mean atomic weight of plasma and, therefore, slower shock velocity (for the same upstream temperature and the shock Mach number), reducing the non-adiabatic heating of ions. Conversely, at later times ($\tau>10^{12}\,\rm cm^{-3}\,s$), the presence of He ions keeps the Fe~XXV ions slightly hotter, because He ions have the longest equilibration time scale.

\end{appendix}

\label{lastpage}
\end{document}